# Permutable SOS (Symmetry Operational Similarity)

**Sang-Wook Cheong, Seongjoon Lim, Kai Du, and Fei-Ting Huang**

Rutgers Center for Emergent Materials and Department of Physics and Astronomy, Piscataway, NJ 08854, USA

Correspondence: S.-W. Cheong (email: sangc@physics.rutgers.edu)

## ABSTRACT

**Based on symmetry consideration, quasi-one-dimensional (1D) objects, relevant to numerous observables or phenomena, can be classified into eight different types. We provide various examples of each 1D type, and discuss their Symmetry Operational Similarity (SOS) relationships, which are often permutable. A number of recent experimental observations, including current-induced magnetization in polar or chiral conductors, non-linear Hall effect in polar conductors, spin-polarization of tunneling current to chiral conductors, and ferro-rotational domain imaging with linear gyration are discussed in terms of (permutable) SOS. In addition, based on (permutable) SOS, we predict a large number of new phenomena in low symmetry materials that can be experimentally verified in the future.**

**Symmetry defines evident beauty.**
**Once broken, it brings observables.**
**Unveiling duality and permutability.**
**These tricks that Nature plays become the crystal ball to see its hidden beauty.**
**In Nature, nothing is perfect, yet everything is perfect*.**

**INTRODUCTION**

The concept of SOS (symmetry operational similarity)[1,2] between specimen constituents (i.e., periodic crystallographic or magnetic lattices in external fields or other environments, etc., and also their time evolution) and measuring probes/quantities (observables such as propagating light, electrons or other particles in various polarization states, including light or electrons with spin or orbital angular momentum, bulk polarization or magnetization, etc., and also experimental setups to measure, e.g., Hall-type effects) in relation to broken symmetries is powerful to understand and predict observable physical phenomena in low symmetry materials.[3] This SOS relationship includes when specimen constituents have more, but not less, broken symmetries than measuring probes/quantities do. In other words, in order to have a SOS relationship, specimen constituents cannot have higher symmetries than measuring probes/quantities do. The power of the SOS approach lies in providing simple and physically transparent views of otherwise complicated and unintuitive phenomena in complex materials.

In conjunction with the earlier publications,[1-3] we here discuss (quasi-)one-dimensional objects (simply referred as 1D objects): objects can be specimen constituents (periodic lattices in the presence of external perturbations) or measuring probes/quantities (observables), and (quasi-)1D refers objects that has a certain rotation symmetry around the 1D direction, and all rotational symmetry, possibly except $C_2$ symmetry, is broken around any directions perpendicular to the 1D direction. Note that periodic crystallographic or magnetic lattices can have 2-, 3-, 4-, or 6-fold rotational symmetries ($C_2$, $C_3$, $C_4$, or $C_6$ symmetries in the standard crystallographic notations, respectively), 1D specimen constituents must have at least, $C_2$ or $C_3$ symmetry around the 1D direction, and broken all of $C_3$, $C_4$, and $C_6$ symmetries around any directions perpendicular to the 1D direction. One example of 1D objects is cycloidal spins having SOS with polarization (***P***), and cycloidal spins do have $C_2$ symmetry around ***P***, but no rotation symmetry, including $C_2$, around any direction perpendicular to the ***P*** direction.[1]

Here, we define **R**=π (i.e. 2-fold) rotation operation with the rotation axis perpendicular to the 1D object direction, <u>**R**</u>=π (i.e. 2-fold) rotation operation with the rotation axis along the 1D object direction, **I**=space inversion, **M**=mirror operation with the mirror perpendicular to the 1D object direction, <u>**M**</u>= mirror operation with the mirror plane containing the 1D object direction, **T**=time reversal operation. (In the standard crystallographic notations, "**R**, <u>**R**</u>, **I**, **M**, <u>**M**</u>, and **T**" are "$C_{2\perp}$, $C_{2\parallel}$, $\bar{1}$, $m_{\perp}$, $m_{\parallel}$, 1′, respectively.[1] Our notations are intuitive to consider 1D objects. In some cases, we specify 2-fold rotation and mirror operations further using the notations of **R•**, **R_**, **R|**, **M□**, **M|**, **and M_** with the relevant axes/planes, defined with respect to the page plane, in the subscriptions. Note that since we consider the 1D objects, translational symmetry is ignored. For example, **T** symmetry in a simple 1D antiferromagnet is not broken since we ignore translation.

Hlinka has proposed that there exist eight kinds of vectorlike physical quantities in term of their invariance under spatiotemporal symmetry operations.[4] In fact, our 1D objects are exactly like the vectorlike physical quantities in terms of symmetry, so there are eight kinds of 1D objects. We will call them as "𝓓, 𝓒, 𝓟, 𝓐, 𝓓′, 𝓒′, 𝓟′, 𝓐′". 𝓓, 𝓒, 𝓟, and 𝓐 refer to Director, Chirality, Polarization, and rotational Axial (electric toroidal) vector, respectively, and ′ refers to the time reversal broken version. All left-hand-side 1D objects (𝓓, 𝓒, 𝓓′, 𝓒′) in Fig. 1 have not-broken **R** (i.e. they are director-like) and all right-hand-side 1D objects (𝓟, 𝓡, 𝓟′, 𝓡′) have broken **R** (i.e. they are vector-like).

## RESULTS AND DISCUSION

We have attempted to identify various exemplary 1D objects that can be classified into each of the eight Hlinka's classifications[4], which are listed in Fig. 1 and described below:

$\mathcal{D}$ (no broken symmetry): Director. Simple antiferromagnetic order in a chain, up-up-down-down antiferromagnetic order in a chain, and up-up-down-down antiferromagnetic order in a chain with two kinds of bonding.

$\mathcal{C}$ (broken {**I**,**M**,**<u>M</u>**}): Chirality. Crystallographic chirality, helical spins, magnetic toroidal moments with out-of-plane canted moment, linear momentum (velocity) with spin or orbital angular momentum, circularly-polarized light, and vortex beam having orbital angular momentum.

$\mathcal{P}$ (broken {**R**,**I**,**M**}): Polarization (***P***). Strain gradient, thermal gradient, and cycloidal spins.

$\mathcal{A}$ (broken {**R**,**<u>M</u>**}): Axial vector of ferro-rotation. Crystallographic rotational distortions, and electric toroidal moment.

$\mathcal{D}'$ (broken {**T**}): Director with broken {**T**}. Up-up-down-down antiferromagnetic order in a chain with alternating two kinds of chiral bonding environments. This corresponds to the ground magnetic state of α-$Fe_2O_3$ below the Morin transition, at which canted magnetic moment disappears.

$\mathcal{C}'$ ((broken {**I**,**M**,**<u>M</u>**,**T**}): Chirality with broken {**T**}. Simple antiferromagnetic order in a chain with two kinds of bonding, simple antiferromagnetic order in a chain with alternating two kinds of chiral bonding environments, which corresponds to the magnetic sate of $Cr_2O_3$, magnetic monopole, and an in-plane antiferromagnetic order on honeycomb lattice.

$\boldsymbol{\mathcal{P}}'$ (broken {**R**,**I**,**M**,**T**}): Velocity (***k***, linear momentum, wave vector). The act of applying electric field inducing electric current, magnetic toroidal moments, an in-plane antiferromagnetic order on a honeycomb lattice, Ising antiferromagnetic order on a honeycomb lattice, and antiferromagnetic order in a chain with alternating two kinds of chiral bonding environments, corresponding to the spin-flopped state of $Cr_2O_3$. Note that all specimen constituents with the $\boldsymbol{\mathcal{P}}'$ character can exhibit non-reciprocal directional dichroism.

$\boldsymbol{\mathcal{A}}'$ (broken {**R**,$\underline{\mathbf{M}}$,**T**}): Axial vector with broken {**T**}. Magnetization (***M***), and two kinds of magnetic order in kagome lattice having SOS with M.

As exemplary cases, we, first, discuss how our SOS approach can be helpful to understand the physics of $Cr_2O_3$ with linear magnetoelectricity and $\alpha$-$Fe_2O_3$ with a canted magnetic moment and also to predict a new phenomenon in $\alpha$-$Fe_2O_3$. In both $Cr_2O_3$ and $\alpha$-$Fe_2O_3$, the 3D lattice consists of a simple hexagonal stacking of 1D c-direction chains of magnetic ions with alternating left-right chiral bonding environments, as shown in Fig. 2(a). Magnetic orders in $Cr_2O_3$ and $\alpha$-$Fe_2O_3$ are schematically shown in Fig. 2(b) and 2(g), respectively, so they correspond to $\boldsymbol{\mathcal{C}}'$ and $\boldsymbol{\mathcal{D}}'$, respectively. In the case of $Cr_2O_3$, the magnetic state in Fig. 2(b) has broken {**I**,**M**,$\underline{\mathbf{M}}$,**T**}, and we can construct a SOS relationship by adding further broken symmetries such as broken {**R**,**I**,**M**} of ***P*** or electric field. As shown in Fig. 2(c) and (d), the magnetic state of $Cr_2O_3$ in the presence of an electric field does have SOS with ***M***, so it shows a diagonal linear magnetoelectric effect.[5] Applying a large magnetic field along the c axis induces a spin-flopped state, shown in Fig. 2(e), and this state exhibits an off-diagonal linear magnetoelectric effect, as the SOS in Fig. 2(e) indicates. The canted magnetic state of α-$Fe_2O_3$ above the Morin transition is consistent with the SOS in Fig. 2(f). The magnetic state of α-$Fe_2O_3$ below the Morin transition, schematically shown in Fig. 2(g), shows no SOS with ***M***, so the canted magnetic moment disappears below the Morin transition. Note that the schematic in Fig. 2(b) has broken {$\underline{\mathbf{M}}$,**M⊗R**,**T**}, so is supposed to exhibit a magnetooptical Kerr effect (MOKE), and indeed, does show non-zero MOKE in $Cr_2O_3$.[6] However, the schematics in Fig. 2(f) and (g) has, for example, not-broken **M**□, so $Fe_2O_3$ does not show any MOKE. It is very interesting to notice that the magnetic state of α-$Fe_2O_3$ below the Morin transition in the presence of an electric field does show SOS with ***k***, which means that it can exhibit a non-reciprocal effect such as an optical-diode effect, which can be experimentally confirmed in the future.

## Permutable SOS of $\mathscr{P}$ x $\mathscr{P}'$ ≈ $\mathscr{A}'$

The usefulness of the SOS approach is that it can be commonly applied to various physical quantities that share the same set of broken symmetries. $\mathscr{P}$ can be polarization (***P***) or an electric field (***E***), but it can be also a temperature gradient, strain gradient, or surface effective electric field, $\mathscr{P}'$ can be velocity/wave vector (***k***) of electric current, spin wave or thermal current, or toroidal moment, and $\mathscr{A}'$ is magnetization (***M***) or magnetic field (***H***). Moreover, the SOS relationships established among them can sometimes be permutable. Fig. 3(a), (b) and (c) shows pictorially the permutable SOS relationships of $\mathscr{P}$ x $\mathscr{A}'$ ≈ $\mathscr{P}'$, $\mathscr{A}'$ x $\mathscr{P}'$ ≈ $\mathscr{P}$, $\mathscr{P}'$ x $\mathscr{P}$ ≈ $\mathscr{A}'$, respectively. Note that all of $\mathscr{P}$, $\mathscr{P}'$, and $\mathscr{A}'$ are vector-like.

It turns out that the motion of quasi-particles such as electrons, spin waves, phonons, and photons in a specimen or the motion of the specimen itself in specimen constituents can be non-reciprocal if the specimen constituents have SOS with $\boldsymbol{k}$. When $+\boldsymbol{k}$ becomes $-\boldsymbol{k}$ under a symmetry operation, while a specimen constituent where quasiparticles are moving with $\pm\boldsymbol{k}$ is invariant under the symmetry operation, then the experimental situation becomes reciprocal. On the other hand, when a specimen constituent has SOS with $\boldsymbol{k}$, then there is no symmetry operation that can connect these two experimental situations: one specimen constituent with $+\boldsymbol{k}$ and the identical specimen constituent with $-\boldsymbol{k}$. Thus, the experimental situation can become non-reciprocal, even though the magnitude of the non-reciprocal effects cannot be predicted. Thus, Fig. 3(a) means that if $\boldsymbol{H}$ is applied to any polar magnets, then electron, light, phonons or spin wave propagation can be non-reciprocal, i.e. exhibit directional dichroism. Indeed, for example, diode-like light propagation has been observed in polar $(Fe,Zn)_2Mo_3O_8$ in $\boldsymbol{H}$ [7], and electron transport in polar BiTeBr [8] in $\boldsymbol{H}$ turns out to be non-reciprocal in experimental configurations consistent with Fig. 3(a). We anticipate that applying in-plane $\boldsymbol{H}$ to polar $Fe(Mn,Ni,Zn)_2Mo_3O_8$ or polar (and chiral) $Ni_3TeO_6$ [9] will induce non-reciprocal spin wave along the in-plane direction perpendicular to $\boldsymbol{H}$.

Figure 3(b) represents all kinds of (anomalous) Hall effects (Hall, Ettingshausen, Nernst, and thermal Hall). Here, ($\boldsymbol{k}$ and $\boldsymbol{P}$) represents (electric current and induced voltage), (electric current and induced thermal gradient), (thermal gradient and induced voltage), and (thermal gradient and induced thermal gradient) for the Hall, Ettingshausen, Nernst, and thermal Hall effects, respectively. Fig. 3(c) can also represent numerous novel physical phenomena, including the off-diagonal linear magnetoelectric effect in $Cr_2O_3$ in the spin-flopped state (Fig. 2(e)), the spin Hall effect, the spin Nernst (or thermal spin Hall) effect, current-induced magnetization in polar conductors and also non-linear Hall effects in polar conductors.[10] When $\boldsymbol{P}$ represents the surface effective electric field, electric current and thermal current for $\boldsymbol{k}$ can induce spin current flow along the direction perpendicular to ***P*** and the current direction – the resulting surface magnetization is represented by ***M***. Here, the electric current and the thermal current correspond to the spin Hall effect and the spin Nernst (or thermal spin Hall) effect, respectively. Fig. 3(c) also represents that electric current, thermal current or light propagation along a direction perpendicular to ***P*** in polar materials can induce ***M*** along the direction perpendicular to both the ***P*** and the current/light propagation directions. Even though ***M*** in polar materials induced by thermal current/light has not been reported, ***M*** in polar conductors induced by electric current has been reported, interestingly in an exotic material of strained monolayer $MoS_2$.[11-12] It is interesting to observe this current-induced ***M*** in bulk non-magnetic polar conductors such as $Ca_3Ru_2O_7$, GeTe and BiTeI(Br,Cl).[13-15] In addition, it will be fundamentally important to observe ***M*** in polar materials induced by thermal current or light – in this case, the polar materials do not have to be electric conductors, in principle.

The monolayer of 2H-$MoS_2$, shown in Fig. 4(a), is non-centrosymmetric, but non-polar due to the presence of $C_3$ symmetry. However, uniaxial uniform strain along a particular in-plane direction breaks the $C_3$ symmetry and induces ***P*** in the direction perpendicular to the strain. When electric current is applied to an in-plane direction perpendicular to the strain-induced ***P***, then ***M*** develops along the out-of-plane direction, which is depicted in Fig. 4(b) and (c). Tensile and compressive strains result in opposite ***M***s. A similar situation, but without external strain, should also work for monolayer $T_d$-$WTe_2$ or $T_d$-$MoTe_2$. Monolayer $T_d$-$WTe_2$ or $T_d$-$MoTe_2$ with W/Mo zigzags does have in-plane polarization perpendicular to the zigzag direction. When electric current is applied along the zigzag direction, ***M*** along the out-of-plane direction should be induced, which requires a future experimental confirmation. In fact, this ***M*** induced by electric current in polar conductors is directly related with the presence of non-linear Hall effect that we will discuss next.

First, we attempt to figure out the requirement for quadratic-type non-linear Hall effect (NLHE) without any applied magnetic field. We consider four Fig. 4(g)-(j) situations of Hall-type voltages with an applied external electric field (+-$\boldsymbol{E}_{ext}$). Any of {**R**|,**M**|,**M**□,**T**} links between Fig. 4(g) and (h) (or (g) or (h) itself). NLHE should not depend on the sign of the external E, so NLHE may require not-broken {**R**|,**M**|,**M**□,**T**}. Now, any cross-link between Fig. 4(g)/(h) and (i)/(j) can happen with {**R•**,**R_**,**I**,**M_**}, so not-broken any of {**R•**,**R_**,**I**,**M_**}, in addition to not-broken {**R**|,**M**|,**M**□,**T**}, will lead to NLHE to vanish. Therefore, broken {**R•**,**R_**,**I**,**M_**}is required to have non-zero NLHE. {**R•**,**R_**,**I**,**M_**} becomes {**R**,**I**,**M**} with the respect to the vertical reference direction on the page plane. Thus, having SOS relationship with ***P*** along the vertical direction is the necessary and sufficient condition for non-zero NLHE. ***P*** has not-broken {**R**|,**M**|,**M**□,**T**}, but in principle, any specimen constituents having SOS with ***P*** can exhibit non-zero NLHE. In other words, the specimen constituents with non-zero NLHE can have lower symmetry than ***P*** (i.e., can have broken some of {**R**|,**M**|,**M**□,**T**}).

Now, there exists a straightforward but elegant relationship between current-induced magnetization with NLHE in polar conductors. When an electric current, ***J***, is applied to a polar conductor along a direction perpendicular to ***P***, ***M*** is induced along a direction perpendicular to both ***J*** and ***P***. This, induced ***M*** acts like a magnetic field for the standard Hall effect, and Hall voltage along the ***P*** direction goes like $J \mathrm{x} M \propto J^2$, which explains quadratic-type non-linear Hall effect in polar conductors. Therefore, strained monolayer $MoS_2$ should exhibit NLHE along the strain-induced ***P*** direction. In fact, in monolayer $WTe_2$, NLHE, completely consistent with the above discussion, has been observed. Note that bulk non-magnetic polar conductors such as $Ca_3Ru_2O_7$, GeTe and BiTeI(Br,Cl) should exhibit NLHE, in addition to current-induced $\boldsymbol{M}$.

We also note that the so-called circular dichroism Hall effect can be understood on a similar footing.[16] Illumination of circularly-polarized light on a chiral material can induce electric current, which is called a circular photogalvanic effect (CPGE).[17-19] In the case of CPGE, one can consider if or not the entire experimental setup with induced current changes in a systematic manner under all symmetry operations – the so-called symmetry operational systematics. It turns out that in terms of symmetry operational systematics, a similar phenomenon can occur in a material with a net magnetic moment, rather than chirality.[1] Now, with the consideration of symmetry operational systematics, one can also show that illumination of circularly-polarized light on any materials can induce $\boldsymbol{M}$. This circularly-polarized light-induced $\boldsymbol{M}$ can act like a magnetic field, so when electric current flows along a direction perpendicular to the induced $\boldsymbol{M}$, then a Hall voltage can develop along the direction perpendicular to both $\boldsymbol{M}$ and current without any applied magnetic field, which is called a circular dichroism Hall effect (CDHE). Therefore, in principle, this CDHE can occur in any materials where electric current flow is possible. This CDHE can be also understood in terms of the SOS relationship in Fig. 4(f). Emphasize that the specimen showing with a blue rectangle appears to have no broken symmetries, but the SOS relationship still hold even if the specimen has any broken symmetries by the definition of a SOS relationship. This again guarantees that CDHE can occur in any materials where electric current flow is possible. In addition, the electric current flow can be replaced by thermal current, and also $\boldsymbol{P}$ can stand for a thermal gradient, rather than a Hall voltage. Therefore, there can exist Ettingshausen, Nernst, or thermal Hall effect versions of CDHE in any materials with circularly-polarized light (or vortex beam with orbital angular momentum) illumination without any applied magnetic field.

## Permutable SOS of $\mathcal{C} \bullet \mathcal{P}' \approx \mathcal{A}'$, $\mathcal{A} \bullet \mathcal{A}' \approx \mathcal{D}'$, and $\mathcal{P} \bullet \mathcal{D}' \approx \mathcal{P}'$

We have also the dot product-type, rather than cross product-type, SOS relationships of $\mathcal{C} \bullet \mathcal{P}' \approx \mathcal{A}'$, $\mathcal{A} \bullet \mathcal{A}' \approx \mathcal{D}'$, and $\mathcal{P} \bullet \mathcal{D}' \approx \mathcal{P}'$, as well as their permutable SOS relationships, summarized below: The broken symmetry operations that do not cancel on the left-hand side are colored in green as the resultant broken symmetries on the right-hand side, while canceled ones are colored in dark gray.

| | | | | |
|---|---|---|---|---|
| [1] $\mathcal{C} \bullet \mathcal{P}'$ | ≈ Broken {I,M,$\underline{\text{M}}$} | ⊗ Broken {R,I,M,T} | = Broken {R,$\underline{\text{M}}$,T} | ≈ $\mathcal{A}'$ |
| [2] $\mathcal{C}' \bullet \mathcal{P}$ | ≈ Broken {I,M,$\underline{\text{M}}$,T} | ⊗ Broken {R,I,M} | = Broken {R,$\underline{\text{M}}$,T} | ≈ $\mathcal{A}'$ |
| [3] $\mathcal{C}' \bullet \mathcal{P}'$ | ≈ Broken {I,M,$\underline{\text{M}}$,T} | ⊗ Broken {R,I,M,T} | = Broken {R,$\underline{\text{M}}$} | ≈ $\mathcal{A}$ |
| [4] $\mathcal{A} \bullet \mathcal{C}'$ | ≈ Broken {R,$\underline{\text{M}}$} | ⊗ Broken {I,M,$\underline{\text{M}}$,T} | = Broken {R,I,M,T} | ≈ $\mathcal{P}'$ |
| [5] $\mathcal{A}' \bullet \mathcal{C}'$ | ≈ Broken {R,$\underline{\text{M}}$,T} | ⊗ Broken {I,M,$\underline{\text{M}}$,T} | = Broken {R,I,M} | ≈ $\mathcal{P}$ |
| [6] $\mathcal{A}' \bullet \mathcal{C}$ | ≈ Broken {R,$\underline{\text{M}}$,T} | ⊗ Broken {I,M,$\underline{\text{M}}$} | = Broken {R,I,M,T} | ≈ $\mathcal{P}'$ |
| [7] $\mathcal{P} \bullet \mathcal{A}'$ | ≈ Broken {R,I,M} | ⊗ Broken {R,$\underline{\text{M}}$,T} | = Broken {I,M,$\underline{\text{M}}$,T} | ≈ $\mathcal{C}'$ |
| [8] $\mathcal{P}' \bullet \mathcal{A}$ | ≈ Broken {R,I,M,T} | ⊗ Broken {R,$\underline{\text{M}}$} | = Broken {I,M,$\underline{\text{M}}$,T} | ≈ $\mathcal{C}'$ |
| [9] $\mathcal{P}' \bullet \mathcal{A}'$ | ≈ Broken {R,I,M,T} | ⊗ Broken {R,$\underline{\text{M}}$,T} | = Broken {I,M,$\underline{\text{M}}$} | ≈ $\mathcal{C}$ |
| [10] $\mathcal{P} \bullet \mathcal{A}$ | ≈ Broken {R,I,M} | ⊗ Broken {R,$\underline{\text{M}}$} | = Broken {I,M,$\underline{\text{M}}$} | ≈ $\mathcal{C}$ |
| [11] $\mathcal{A} \bullet \mathcal{A}'$ | ≈ Broken {R,$\underline{\text{M}}$} | ⊗ Broken {R,$\underline{\text{M}}$,T} | = Broken {T} | ≈ $\mathcal{D}'$ |
| [12] $\mathcal{P} \bullet \mathcal{D}'$ | ≈ Broken {R,I,M} | ⊗ Broken {T} | = Broken {R,I,M,T} | ≈ $\mathcal{P}'$ |

Note that $\mathcal{C}$, $\mathcal{C}'$, $\mathcal{D}$, and $\mathcal{D}'$ are director-like (i.e. not-broken {**R**}), so the resulting SOS orientation is fixed by vector-like objects of $\mathcal{P}$, $\mathcal{A}$, $\mathcal{P}'$, or $\mathcal{A}'$. Thus, the directions of $\mathcal{C}$, $\mathcal{D}'$, and $\mathcal{C}'$ in Fig, 5 can be in the vertical, horizontal, or out-of-plane direction. [1]-[10] list all SOS relationships permutable from $\mathcal{C} \bullet \mathcal{P}' \approx \mathcal{A}'$: each SOS relationship has two [′]s. A similar sets of permutable SOS relationships are possible from [10] - [12]. First, Fig. 2(c)/(d), reproduced in Fig. 5(a), and Fig. 2(h)/(i) correspond to [2] and [12], respectively. When ***E*** and ***M*** in Fig. 2(c)/(d) are replaced by ***H*** and ***P***, respectively, it corresponds to [5]. This demonstrates a permutable SOS relationship as well as the permutable (and duality) nature of linear magnetoelectric effects.

Permutable SOS relationships of [2], [5], and [7] can be also well exemplified in hexagonal rare-earth ferrites[20]. The $A_2$ phase of the system carries ***P*** ($\mathcal{P}$) along the *c* axis and in-plane magnetic monopole ($\mathcal{C}'$), and a net magnetic moment ***M*** ($\mathcal{A}'$) along the *c* axis is, indeed, observed experimentally, which is neatly consistent with [2]. Through a spin reorientation transition, the magnetic state transforms into the $A_1$ phase with magnetic toroidal moments ($\mathcal{P}'$), where the net magnetic moment M ($\mathcal{A}'$) is no longer allowed according to $\mathcal{P} \bullet \mathcal{P}' \approx \mathcal{D}'$. This is another premium example of how these permutable SOS relationships can be used to explain and predict physical properties in real materials.

Figure 5(b) corresponds to [6], which explains longitudinal and also transverse magnetochiral effects, i.e non-reciprocal electronic transport or optical effects in chiral materials in the presence of external magnetic fields.[21] Note that a twisted graphene bilayer is chiral, so exhibits a natural optical activity,[22] and should exhibit non-reciprocal optical and also transport effects in the presence of magnetic fields, which need to be experimentally verified. Their inverse effects in Fig. 5(c), corresponding to [1], include magnetization induced by electric current in chiral materials and spin-polarized tunneling in chiral materials. The magnetization induced by electric current in chiral Te single crystals has been reported in a Faraday-type optical experiment as well as a direct magnetization pulse measurement.[23-24] The chirality-dependent spin-polarization of tunneling current was recently observed in chiral $Co_{1/3}NbS_2$ [25] for the first time (see below).

Chiral materials can exhibit natural optical activity, i.e. the rotation of light polarization of transmitted linearly-polarized light. Thus, the Fig. 5(d) SOS relationship, corresponding to [10], explains the so-called linear-gyration, i.e. the linear change of the light polarization rotation of transmitted linearly-polarized light with varying electric field. In fact, this linear-gyration can be utilized to visualize ferro-rotational domains, since opposite ferro-rotational domains show the opposite linear-gyration effects.[1] This idea was, indeed, confirmed in the visualization of ferro-rotational domains in ilmenite $NiTiO_3$, as shown in Fig. 6(b).[26] There exists a large number of ferro-rotational compounds with point group symmetry of $\bar{1}$, $2/m$, $\bar{3}$, $\bar{4}$, $4/m$, $\bar{6}$, or $6/m$.[26] Examples include $Cu_3Nb_2O_8$ ($P\bar{1}$),[27] $RbFe(MoO_4)_2$ ($P\bar{3}$ below 190 K),[28] $CaMn_7O_{12}$ ($R\bar{3}$ below 440 K),[29] $NbO_2$ ($I4_1/a$ below 1081 K),[30] and $Pb_2CoTeO_6$ ($R\bar{3}$ below 370 K).[31] Ferro-rotational domain visualization of these compounds need to be further investigated with the linear-gyration technique.

Figure 6(a) exhibits the atomic model of a topological vortex surrounded by six structural chiral domain boundaries in $Co_{1/3}NbS_2$. Since Co intercalants occupy one of three possible Nb-atop sites, A, B, and C, there exist six possible combinations considering the two layers within a unit cell, exhibiting A/B-type stacking. Solid and dashed lines refer to structural chiral domain boundaries in the first and the second layer, respectively. As a result of intercalation site change, the chiral distortion of S layer reverses as depicted by red and blue triangular arrows and gives the alternating structural chirality across the vortex domain boundaries. In reference to the SOS relationship shown in Fig. 5(c), flowing electric current across the chiral structure induces a magnetization along the current direction, which should work even for local probes such as scanning tunneling microscopy (STM). Fig. 6(b) shows the spin-polarized STM (SP-STM) measurement around a topological vortex that clearly reveals the alternating domain contrast locked to the chirality of each domain.[25] It is noteworthy that the induced domain contrast does not require long range ordered magnetic moments, in contrast to the conventional SP-STM. Instead, the specimen constituent comprised of the chiral structure and the tunneling current gives the same set of broken symmetries as magnetization, as shown in Fig. 5(c).

We would like to note finally that ferro-rotation is rather common in numerous materials,[27-31] but does not break inversion symmetry, so difficult to be coupled with measuring probes, and thus has been studied only in a limited degree. The traditional way to study ferro-rotational domains is atomic-resolved transmission electron microscopy (TEM) such as high-resolution TEM or high-angle annular dark-field (HAADF)-STEM to observe the ferro-rotational distortions or ferro-rotational domains[26]. Fig. 6(e) depicts a pair of ferro-rotational domains, corresponding to a clockwise and counterclockwise rotation of the $FeO_6$ octahedra in $RbFe(MoO_4)_2$. Utilizing cryogenic dark-field TEM technique (Fig. 6f), the contrast difference between the two ferro-rotational domains shows up as the brighter $G_a$ spot for the left domain and the darker $G_b$ spot for the right domain are collected. Note that the Bragg spot $G_a$ is converted into $G_b$ in the reciprocal space by the two-fold rotation as two ferro-rotation domains do in real space.[32] The SOS relations of ferro-rotation such as Fig. 5(d) and (e) can be a good guidance for the further exploration of ferro-rotations in the future.

## CONCLUSION

The recent exotic experimental results such as electric current-induced magnetization in strained monolayer $MoS_2$, NLHE in monolayer $WTe_2$, spin-polarized tunneling in chiral $Co_{1/3}NbS_2$ in a paramagnetic state, and non-reciprocal electronic transport in polar BiTeBr can be understood in terms of permutable SOS relationships of 1D objects. We have presented all possible permutable SOS relationships among eight kinds of 1D objects. Based on the permutable SOS relationship, we have proposed a large number of new phenomena, which can be experimentally verified in the future. New predictions include (but are not limited to): [1] a twisted graphene bilayer in the presence of magnetic fields should exhibit non-reciprocal transport and optical effects, [2] NLHE in strained monolayer $MoS_2$, [3] electric current-induced magnetization in monolayer $Mo(W)Te_2$, [4] electric current-induced magnetization, NLHE, and non-reciprocal electronic transport in bulk polar conductors such as $Ca_3Ru_2O_7$, GeTe, and BiTeI(Br,Cl), [5] light- or thermal current-induced magnetization in numerous polar materials, [6] electric/thermal current-, or light-induced magnetization of polar materials at polar domain walls, [7] CDHE in any materials with enough electric conduction, [8] Ettingshausen, Nernst, or thermal Hall effect versions of CDHE in any materials with circularly-polarized light (or vortex beam with orbital angular momentum) illumination without any applied magnetic field, [9] non-reciprocal spin wave propagation in polar materials such as $Fe(Mn,Ni,Zn)_2Mo_3O_8$ or $Ni_3TeO_6$ in magnetic fields, [10] non-reciprocal directional dichroism in $\alpha$-$Fe_2O_3$ below the Morin transition in the presence of an electric field, [11] visualization of ferro-rotational domains of ferro-rotational compounds such as $RbFe(MoO_4)_2$, $CaMn_7O_{12}$, $Cu_3Nb_2O_8$, $NbO_2$, and $Pb_2CoTeO_6$ using a linear-gyration technique, and [12] non-reciprocal directional dichroism of electric current, spin wave or light in ferro-rotational systems in the presence of both electric and magnetic fields. It will be exuberating to confirm these predictions in bulk quantum materials as well as 2D materials in the coming years.

## METHODS

The relationships between specimen constituents (i.e., lattice distortions or spin arrangements in external fields or other environments, etc., and also their time evolution) and measuring probes/quantities (i.e., propagating light, electrons, or other particles in various polarization states, including light or electrons with spin or orbital angular momentum, bulk polarization or magnetization, etc., and also experimental setups to measure, e.g., Hall-type effects) are analyzed in terms of the characteristics under various symmetry operations of rotation, space inversion, mirror reflection, and time reversal. When specimen constituents and measuring probes/quantities share the same broken symmetries, except translation symmetry, they are said to exhibit symmetry operation similarities (SOS), and the corresponding phenomena can occur. In terms of SOS, we have considered the specimen constituents for light- or current-induced magnetization, NLHE, spin polarization of electric current in non-magnetic or paramagnetic states, linear magneto-electric effects, and natural optical activity or non-reciprocal effects in the presence of an electric field. In addition, we also discuss the requirements for the observation of MOKE, Faraday-type effects, and/or anomalous Hall-type effects in terms of broken symmetries.

**DATA AVAILABILITY:** The datasets generated during the current study are available from the corresponding author on reasonable request.

**ACKNOWLEDGEMENTS:** This work was supported by the DOE under Grant No. DOE: DE-FG02-07ER46382 and the NSF under Grant No. DMR-1629059. We have greatly benefited from discussions with Sobhit Singh and David Vanderbilt.

## AUTHOR CONTRIBUTIONS

S.W.C. conceived and supervised the project. All authors were involved in various symmetry analyses. S.L. and F.-T. H wrote the STM and TEM parts, respectively, K. D. wrote some of the example parts and S. W. C. wrote the remaining part.

**COMPETING INTESTES:** The authors declare no conflict of interest.

* Walker A. In nature, nothing is perfect and everything is perfect. Trees can be contorted, bent in weird ways, and they're still beautiful.

## REFERENCES

[1] Cheong, S.-W. SOS: symmetry operational similarity. *npj Quant. Mater*. **4**, 53 (2019).

[2] Cheong, S.-W. Trompe L'oeil Ferromagnetism *npj Quant. Mater.* **5**, 37 (2020).

[3] Cheong, S.-W., Talbayev, D., Kiryukhin, V. & Saxena, A. Broken symmetries, nonreciprocity, and multiferroicity. *npj Quantum Mater.* **3**, 19 (2018).

[4] Hlinka J. Eight types of symmetrically distinct vectorlike physical quantities. *Phys. Rev. Lett.* **113**, 165502 (2014).

[5] Popov, Y. F., Kadomtseva, A. M., Belov, D. V. & Vorob'ev, G. P. Magnetic-field-induced toroidal moment in the magnetoelectric $Cr_2O_3$. *JETP Lett.* **69**, 330–335 (1999).

[6] Krichevtsov, B. B., Pavlov, V. V., Pisarev, R. V. & Gridnev, V. N. Spontaneous nonreciprocal reflection of light from antiferromagnetic $Cr_2O_3$. *J. Phys. Condens. Matter* **5**, 8233–8244 (1993).

[7] Yu, S. *et al.* High-temperature terahertz optical diode effect without magnetic order in polar $FeZnMo_3O_8$. *Phys. Rev. Lett.* **120**, 037601 (2018).

[8] Ideue, T. *et al.* Bulk rectification effect in a polar semiconductor. *Nat. Phys.* **13**, 578-583 (2017).

[9] Yokosuk, M. *et al.* M. Nonreciprocal directional dichroism of a chiral magnet in the visible range. *npj Quant. Mater.* **5**, 20 (2020)

[10] Ma, Q. *et al.* Observation of the nonlinear Hall effect under time-reversal-symmetric conditions. *Nature* **565**, 337–342 (2019).

[11] Lee, J., Wang, Z., Xie, H., Mak, K. F. & Shan, J. Valley magnetoelectricity in single-layer $MoS_2$. *Nat. Mat.* **16**, 887-891 (2017).

[12] Son, J., Kim, K.-H., Ahn, Y. H., Lee, H.-W. & Lee, J. Strain engineering of the Berry curvature dipole and valley magnetization in monolayer $MoS_2$. *Phys. Rev. Lett.* **123**, 036806 (2019).

[13] Yoshida, Y. *et al.* Crystal and magnetic structure of $Ca_3Ru_2O_7$. *Phys. Rev. B* **72**, 054412 (2005).

[14] Rabe, K. M. & Joannopoulos, J. D. Theory of the structural phase transition of GeTe. *Phys. Rev. B* **36**, 6631 (1987).

[15] Ishizaka, K. *et al.* Giant Rashba-type spin splitting in bulk BiTeI. *Nat. Mater.* **10**, 521–526 (2011).

[16] Zhou, Y. & Liu, F. Realization of an antiferromagnetic superatomic graphene: dirac Mott insulator and circular dichroism Hall effect. Nano Lett. In print (2020).

[17] Ivchenko, E. L. & Pikus, G. E. New photogalvanic effect in gyrotropic crystals. *Pis'ma Zh. Eksp. Teor. Fiz.* **27**, 640 (1978); *JETP Lett.* **27**, 604–608 (1978).

[18] Belinicher, V. I. Space-oscillating photocurrent in crystals without symmetry center. *Phys. Lett. A* **66**, 213–214 (1978).

[19] Asnin, V. M. *et al* Observation of a photo-emf that depend on the sign of the circular polarization of the light. *Pis'ma Zh. Eksp. Teor. Fiz*. **28**, 80–84 (1978); JETPLett. 28, 74–77 (1978).

[20] Du, K. *et al.* Vortex ferroelectric domains, large-loop weak ferromagnetic domains, and their decoupling in hexagonal (Lu, Sc)$FeO_3$. *npj Quant. Mater.* **3**, 33 (2018).

[21] Nakagawa, N. *et al.* Magneto-chiral dichroism of $CsCuCl_3$. *Phys. Rev. B* **96**, 121102(R) (2017).

[22] Kim, C. J. *et al.* Chiral atomically thin films. *Nat. Nanotech.* **11**, 520–524 (2016).

[23] Vorobev, L. E. *et al.* Optical-activity in tellurium induced by a current. *JETP Lett*. **29**, 441–445 (1979).

[24] Furukawa, T., Shimokawa, Y., Kobayashi, K. & Itou, T. Observation of current induced bulk magnetization in elemental tellurium. *Nat. Commun.* **8**, 954 (2017).

[25] Lim, S. J., Huang, F.-T., Pan, S., Wang, K. F., Kim, J.-W. & Cheong, S.-W. Magnetochiral spin-polarized tunneling in a paramagnetic state, to be published.

[26] Hayashida T. *et al.* Visualization of ferroaxial domains in an order-disorder type ferroaxial crystal. *Nat. Commun.* **11**, 4582 (2020).

[27] Johnson, R. D. *et al.* $Cu_3Nb_2O_8$: A multiferroic with chiral coupling to the crystal structure. *Phys. Rev. Lett.* **107**, 137205 (2011).

[28] Jin, W. *et al.* Observation of a ferro-rotational order coupled with second-order nonlinear optical fields. *Nat. Phys.* **16**, 42-46 (2020).

[29] Johnson, R. D. *et al.* Giant improper ferroelectricity in the ferroaxial magnet $CaMn_7O_{12}$. *Phys. Rev. Lett.* **108**, 067201 (2012).

[30] Pynn, R., Axe, J. D. & Thomas, R. Structural distortions in the low-temperature phase of $NbO_2$. *Phys. Rev. B* **13**, 2965-2975 (1976).

[31] Ivanov, S. A., Nordblad, P., Mathieu, R., Tellgren, R. & Ritterd, C. Structural and magnetic properties of the ordered perovskite $Pb_2CoTeO_6$. *Dalton Trans.* **39**, 11136-11148 (2010).

[32] Huang, F.-T., Admasu, A. S., Han, M. G. & Cheong, S.-W. TEM on ferro-rotational domains in $RbFeMo_2O_8$. To be published.

## FIGURE LEGENDS

**Figure 1 Eight types of 1D objects.** Each type has a particular set of broken symmetries: $\mathcal{D}$; no broken symmetry, $\mathcal{C}$; broken {**I**,**M**,$\underline{\mathbf{M}}$}, $\mathcal{D}'$; broken {**T**}, $\mathcal{C}'$; broken {**I**,**M**,$\underline{\mathbf{M}}$,**T**}, $\mathcal{P}$; broken {**R**,**I**,**M**}, $\mathcal{A}$; broken {**R**,$\underline{\mathbf{M}}$}, $\mathcal{P}'$; broken {**R**,**I**,**M**,**T**}, $\mathcal{A}'$; broken {**R**,$\underline{\mathbf{M}}$,**T**}. When {**T**} is additionally broken, then $\mathcal{D}$, $\mathcal{C}$, $\mathcal{P}$, and $\mathcal{A}$ become $\mathcal{D}'$, $\mathcal{C}'$, $\mathcal{P}'$, $\mathcal{A}'$, respectively. Note that all left-hand-side 1D objects ($\mathcal{D}$, $\mathcal{C}$, $\mathcal{D}'$, $\mathcal{C}'$) have not-broken {**R**}, so they are director-like, and all right-hand-side 1D objects ($\mathcal{P}$, $\mathcal{R}$, $\mathcal{P}'$, $\mathcal{R}'$) have broken {**R**}, so they are vector-like. A large number of quasi-1D schematic examples are also shown: blue arrows are spins or magnetization, red arrows are polarizations or electric fields, wiggly arrows are velocity vectors or linear momenta, "***+l***" represents angular momentum or circular polarization, light-blue symbols depict the reference 1D direction, green dots are inversion centers, dashed black circles are anions, and golden springs represent mono-axial chiral objects. The reference 1D direction is the horizontal direction on the page plane, except when there is an explicit light-blue symbol depicting the 1D direction.

**Figure 2 Various magnetic states of a representative *c*-direction chain of $Cr_2O_3$ or α-$Fe_2O_3$, and their SOS relationships with *M* or *k* in the presence of electric fields.** (a) Crystallographic structure of a *c*-direction chain of $Cr_2O_3$ or α-$Fe_2O_3$, and left-chiral and right-chiral arrangements of three oxygen tringles are shown in the left-hand-side and right-hand-side projections along the c axis, respectively. (b) The magnetic state of $Cr_2O_3$ with broken {**M**,**M**⊗**R**,**T**}, so can exhibit MOKE. (c) & (d) The magnetic state of $Cr_2O_3$ in ***E*** shows SOS with ***M***. (e) The spin-flopped state of $Cr_2O_3$ in ***E*** shows SOS with ***M***. (f) The magnetic state of α-$Fe_2O_3$ above the Morin transition has SOS with ***M***. (g) The magnetic state of α-$Fe_2O_3$ below the Morin transition has no SOS with any ***M***. (h) & (i) The magnetic state of α-$Fe_2O_3$ below the Morin transition in ***E*** has SOS with ***k***.

**Figure 3 Three permutable SOS among $\mathcal{A}'$, $\mathcal{P}$, and $\mathcal{P}'$.** Magnetization (***M***), polarization (***P***), and velocity (***k***) are representative for $\mathcal{A}'$, $\mathcal{P}$, and $\mathcal{P}'$, respectively. (a) ***P*** x ***M*** has broken {**R**,**I**,**M**,**T**} along the horizontal direction in the page plane, so has SOS with ***k***. (b) ***M*** x ***k*** has broken {**R**,**I**,**M**} along the horizontal direction, so has SOS with ***P***. ***k*** x ***P*** has broken {**R**,$\underline{\mathbf{M}}$,**T**} along the horizontal direction, so has SOS with ***M***.

**Figure 4 Current-induced *M*, circular dichroism Hall effect (CDHE), and non-linear Hall effect (NLHE).** (a)-(c) Monolayer of 2H-$MoS_2$: (a); no strain, (b); uni-directional uniform tensile strain along the vertical direction on the page plane, and (c); uni-directional uniform compressive strain along the vertical direction on the page plane. (d) and (e) show monolayer 1T-Mo(W)$Te_2$ and Td-Mo(W)$Te_2$, respectively. Closed black circles, open black circles, wiggly arrows, red arrows, and blue symbols represent cations, anions, uniform electric currents, polarizations, and induced magnetizations, respectively. Thus, (b), (c) and (e) represent the current-induced magnetization in polar conductors. This current-induced magnetization is directly related with the presence of non-linear Hall effects in polar conductors without any external magnetic field. (f) depicts the SOS relationship for CDHE with electric current flow with ***k***, circular-light illumination with -***k*** and +***l***, and ***P*** representing the induced Hall voltage. (g)-(j) Four possible configurations of Hall-type voltages induced by electric currents in external electric fields.

**Figure 5 Various SOS relationships.** Yin-yang symbol represents ferro-rotation. (a)-(e) correspond to $\mathcal{C}' \bullet \mathcal{P} \approx \mathcal{A}'$, $\mathcal{A}' \bullet \mathcal{C} \approx \mathcal{P}'$, $\mathcal{C} \bullet \mathcal{P}' \approx \mathcal{A}'$, $\mathcal{P} \bullet \mathcal{A} \approx \mathcal{C}$, and $\mathcal{A}' \bullet \mathcal{C} \approx \mathcal{P} \bullet \mathcal{D}' \approx \mathcal{P}'$, respectively.

**Figure 6 Visualization of domains associated with various broken symmetries.** (a) A $Z_6$ vortex of $Co_{1/3}NbS_2$ with chiral domains. (b) Chirality-dependent SP-STM unveils a $Z_6$ vortex.[25] (c) The atomic structure of ferro-rotational domains of $NiTiO_3$.[26] (d) Ferro-rotational domains in $NiTiO_3$ visualized with an electro-gyration technique. (e) The atomic structure of ferro-rotational domains of $RbFeMo_2O_8$. (f) Ferro-rotational domains observed in TEM on $RbFeMo_2O_8$.[32]

Cheong *et al.* Fig. 1

# Eight types of 1D objects

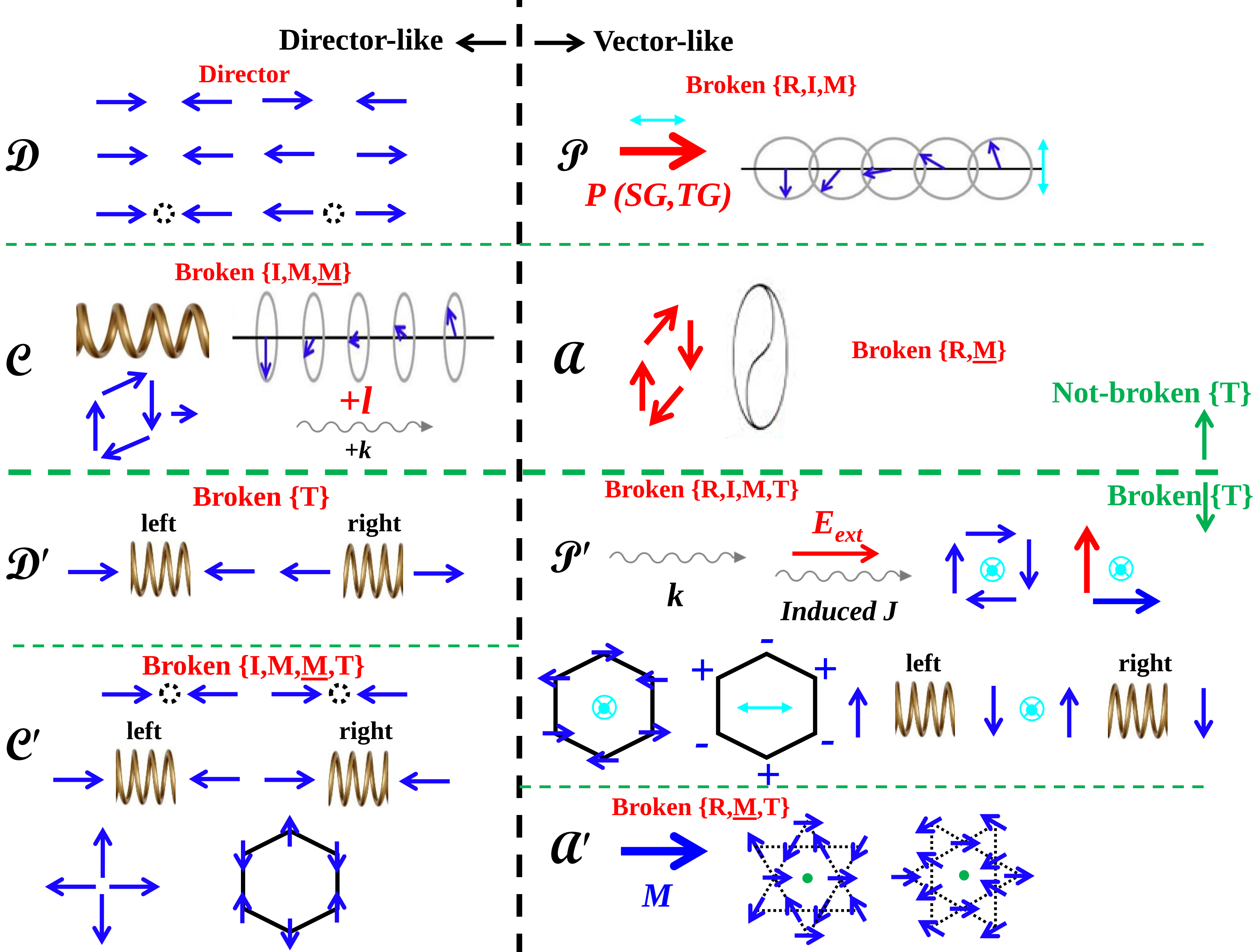

Cheong *et al.* Fig. 2

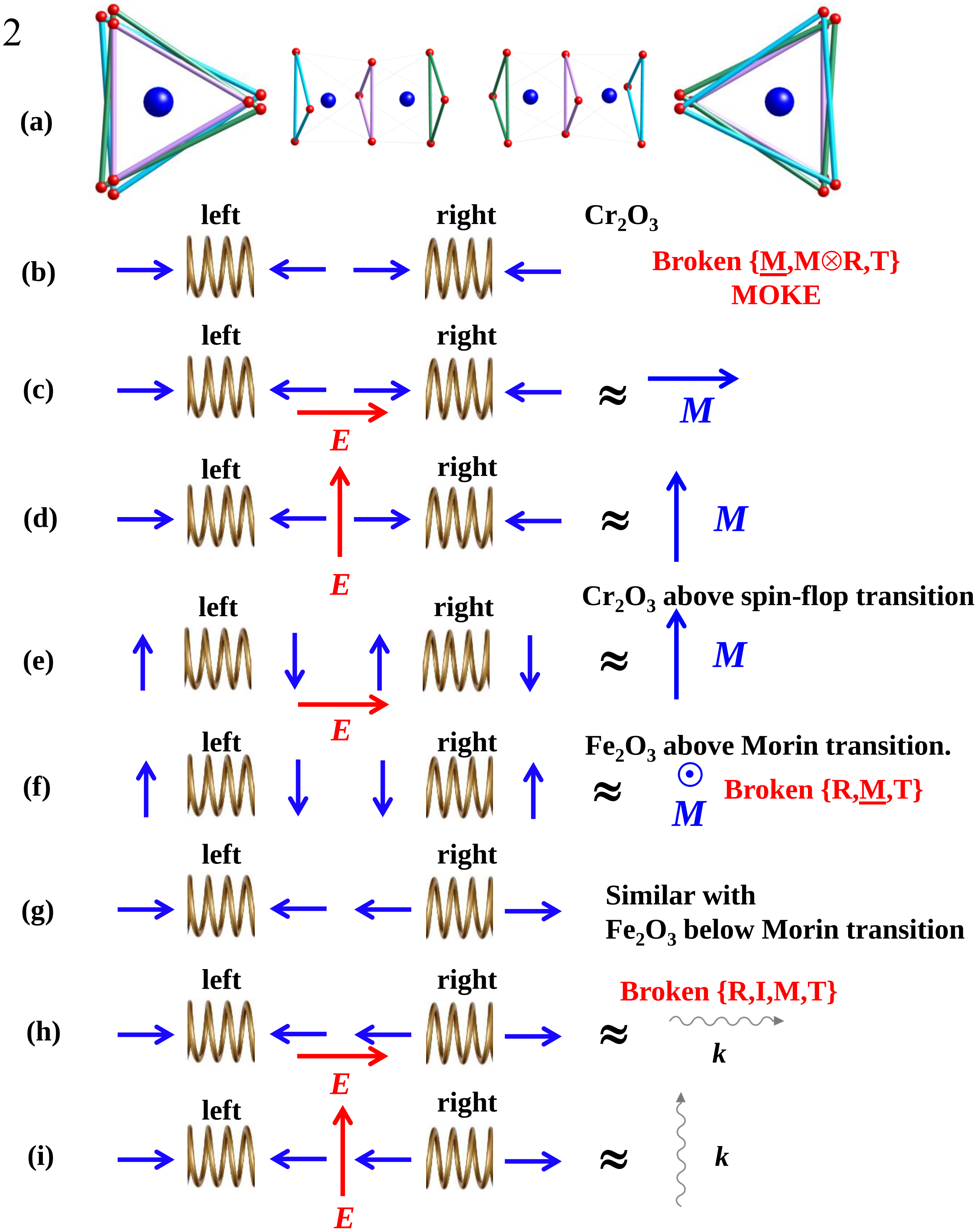

Cheong *et al.* Fig. 3

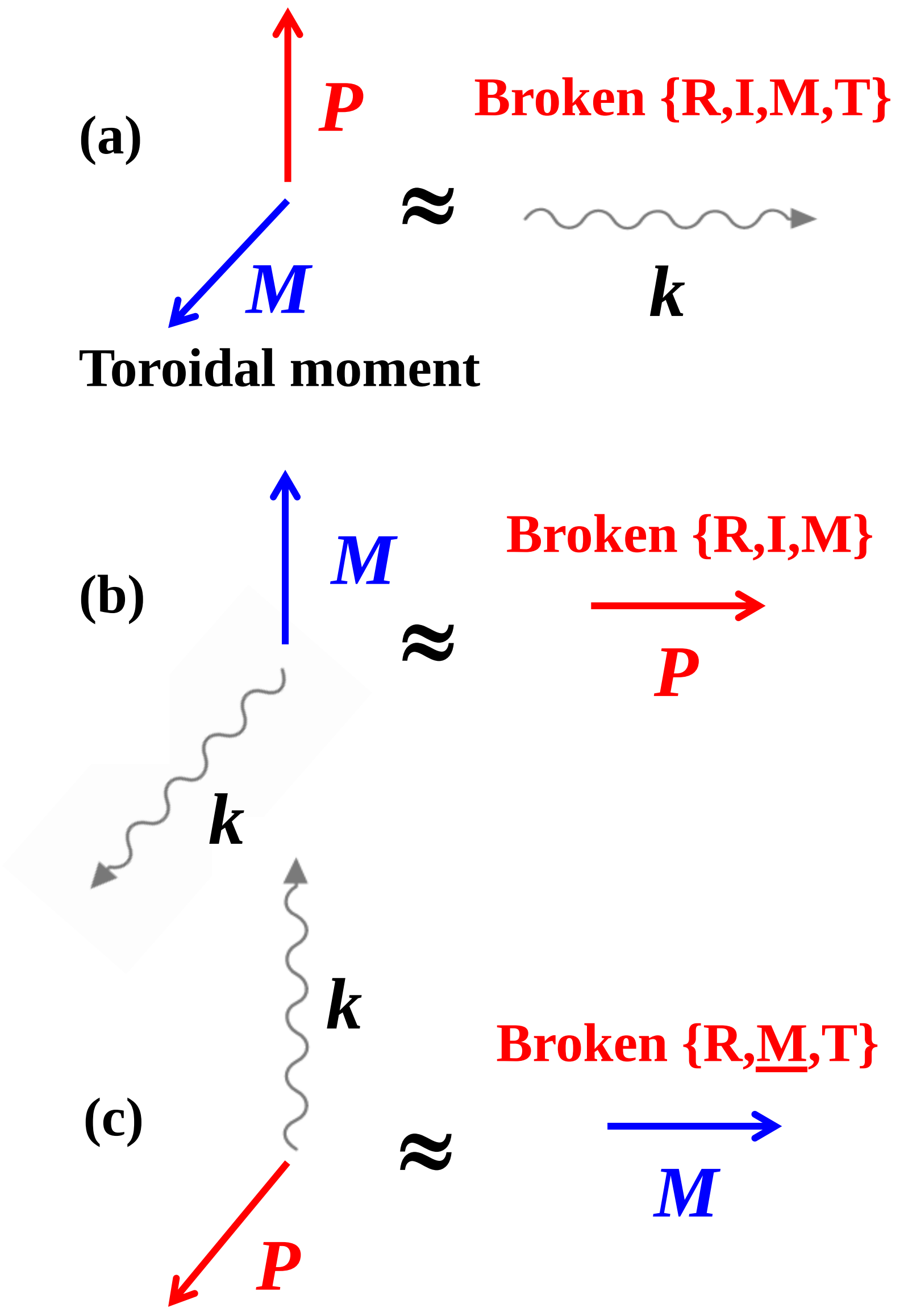

Cheong *et al.* Fig. 4

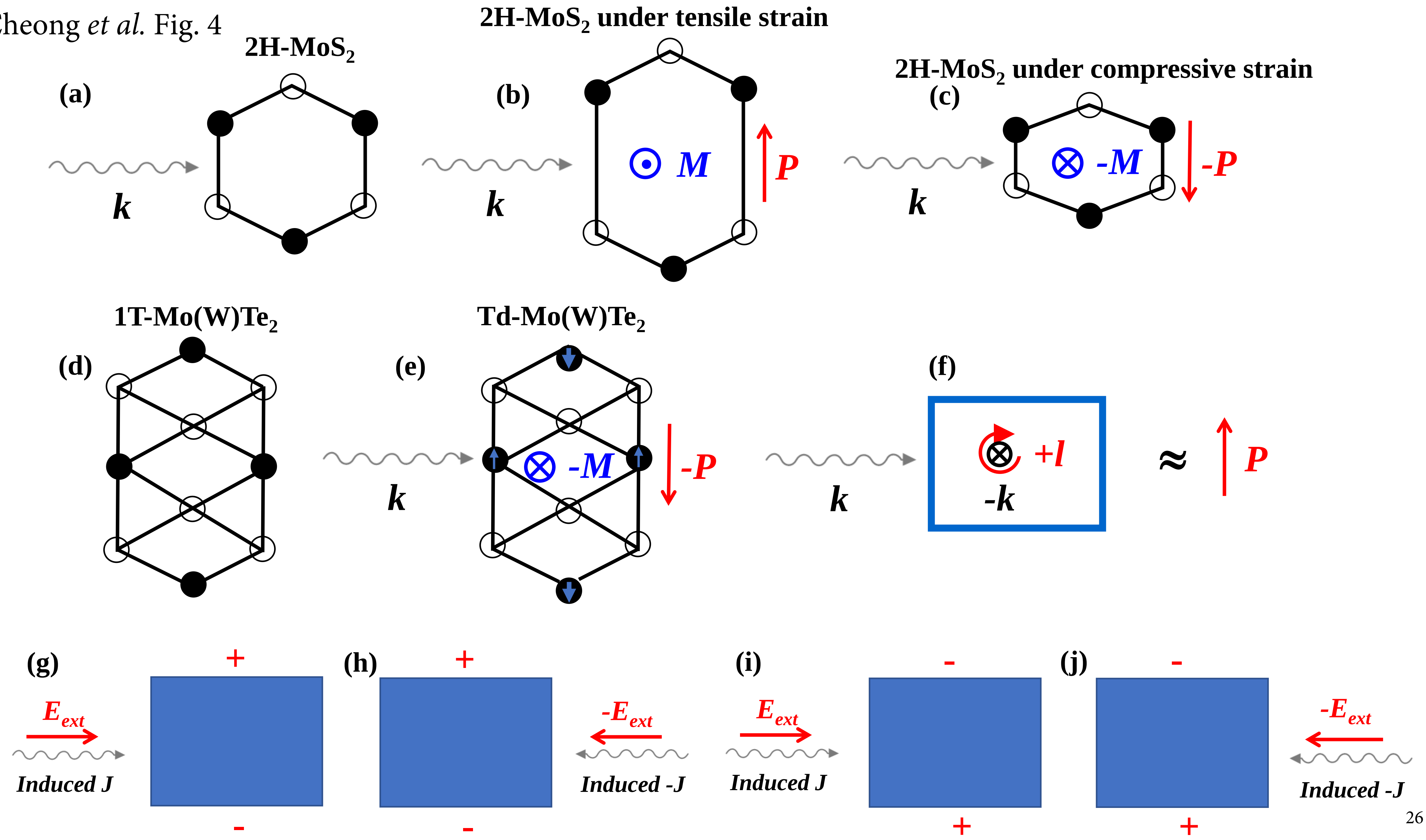

Cheong *et al.* Fig. 5

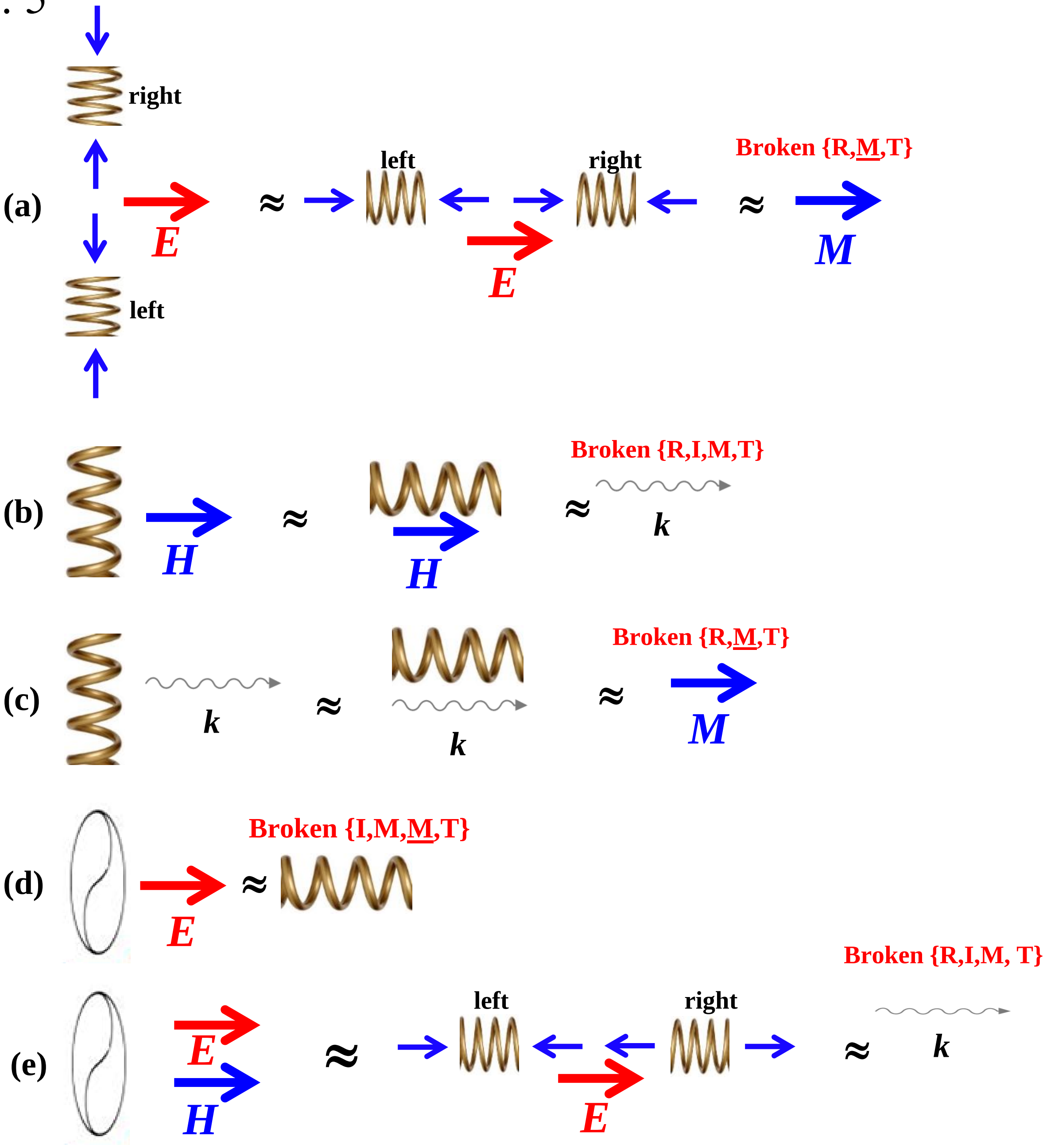

Cheong *et al.* Fig. 6

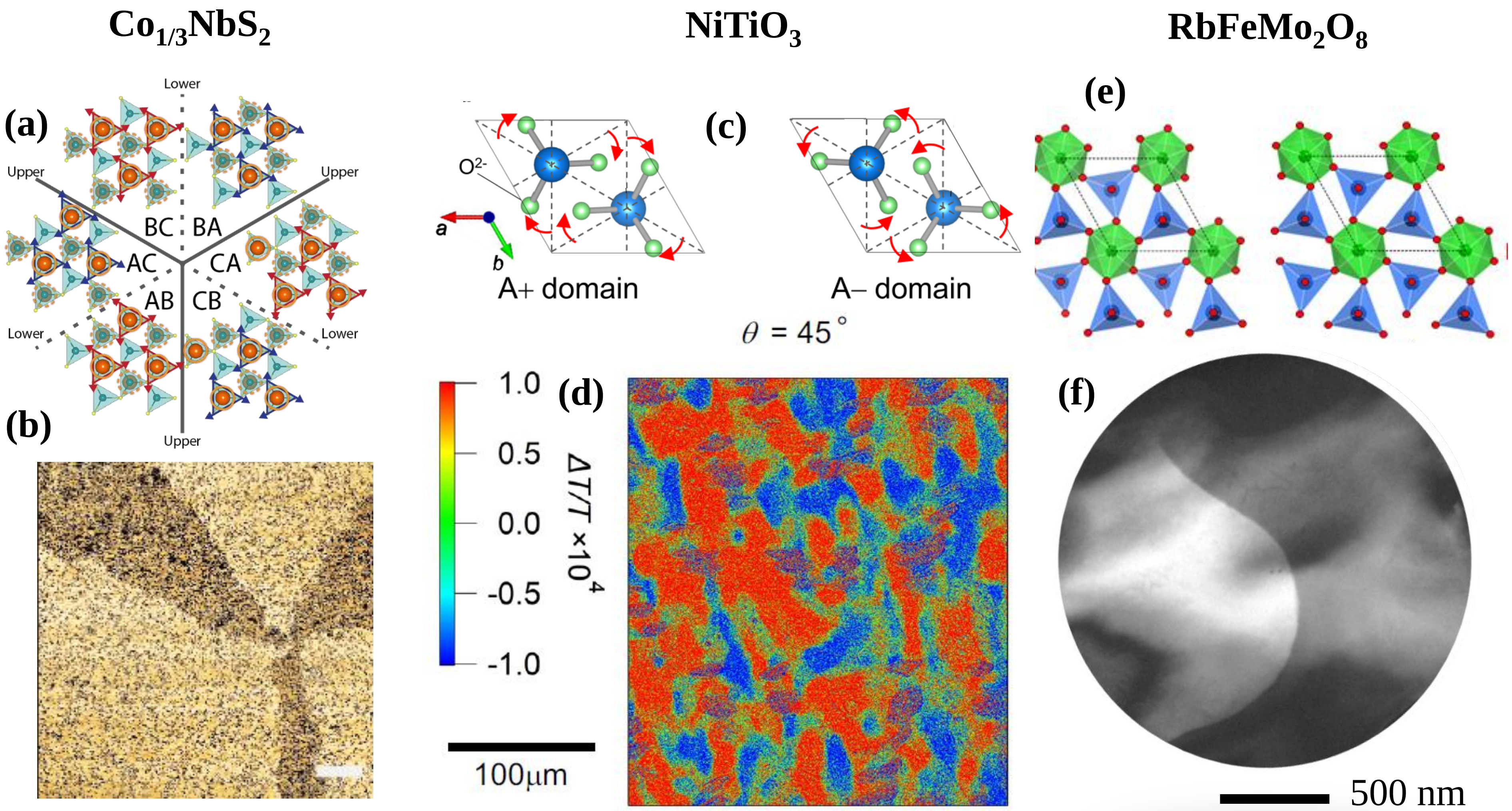